\documentclass[a4paper]{revtex4}
\usepackage{graphicx}
\usepackage{fancyhdr}
\usepackage{amsmath}
\usepackage{color}

\begin{document}

\title{Hierarchy problem, gauge coupling unification at the Planck scale, and vacuum stability
}

%

\author{Yuya Yamaguchi$^{1,2}$}
\affiliation{$^1$ Graduate School of Science and Engineering, Shimane University, Matsue 690-8504, Japan\\
$^2$ Department of Physics, Faculty of Science, Hokkaido University, Sapporo 060-0810, Japan}

\begin{abstract}

To solve the hierarchy problem of the Higgs mass,
 it may be suggested that
 there are no an intermediate scale up to the Planck scale except for the TeV scale.
For this motivation, we investigate possibilities of gauge coupling unification (GCU)
 at the Planck scale ($M_{Pl} = 2.4 \times 10^{18}\,{\rm GeV}$)
 by adding extra particles with the TeV scale mass into the standard model.
We find that
 the GCU at the Planck scale can be realized
 by extra particles including some relevant scalars,
 while it cannot be realized only by extra fermions with the same masses.
On the other hand, when extra fermions have different masses,
 the GCU can be realized around $\sqrt{8 \pi} M_{Pl}$.
By this extension, the vacuum can become stable up to the Planck scale.
\end{abstract}

\maketitle

\thispagestyle{fancy}


\section{Introduction}

The standard model (SM) like Higgs boson has discovered at the LHC experiment,
 and its mass is obtained by the ATLAS and CMS combined experiments as
 $M_h =125.09 \pm 0.21 \ ({\rm stat.}) \pm 0.11 \ ({\rm syst.})\ {\rm GeV}$~\cite{Aad:2015zhl}.
In the SM, this value of the Higgs mass leads the vacuum to be unstable,
 that is, the Higgs quartic coupling $\lambda$ becomes zero below, but close to, the Planck scale
 ($M_{Pl} = 2.435 \times 10^{18}$\,GeV)~\cite{Buttazzo:2013uya}.
The vacuum stability problem may suggest appearance of new physics below the Planck scale.
If new particles appear beyond the SM,
 runnings of the gauge couplings become larger compared to the SM case.
Then, the vacuum can be stable up to the Planck scale,
 since runnings of $\lambda$ also becomes larger. 
Furthermore, the change of the running gauge couplings can realize
 the gauge coupling unification (GCU) at a high energy scale
 (see Ref.~\cite{Giudice:2004tc} for a general discussion).

In addition to the vacuum instability,
 the hierarchy problem of the Higgs mass would be appeared in the SM.
In fact, the quadratic divergence of the Higgs mass term can be always multiplicatively subtracted
 at some energy scale from the Bardeen's argument~\cite{Bardeen:1995kv}.
Since a renormalization group equation (RGE) of the Higgs mass term is proportional to itself in the SM,
 once it is zero at a high energy scale, e.g., the Planck scale,
 it continues to be zero at lower energy scales.
However, if there are heavy particles coupling with the Higgs doublet,
 the Higgs mass term gives logarithmic correction as $M^2 \log (\mu/M)$,
 where $M$ and $\mu$ are mass of the heavy particle and renormalization scale, respectively.
Therefore, the hierarchy problem can be solved
 if no large intermediate scales exist between the electroweak and the Planck scales.

In this paper, we will consider that the Planck scale as a fundamental scale,
 in which the quadratic divergences are assumed to be completely removed out.
To solve the hierarchy problem,
 we do not consider any intermediate scale except for the TeV scale.
Under this context,
 we will investigate possibilities for the realization of the GCU at the Planck scale,
 and discuss the vacuum stability.
This paper is based on our previous work~\cite{Haba:2014oxa}.

\section{Requirement for the GCU} \label{sec:requirement}
We investigate possibilities for the realization of GCU at some high energy scales.
Solving the RGEs,
 we can see the behavior of the gauge couplings in an arbitrary high energy scale.
The one-loop level RGEs of the gauge couplings $\alpha_i=g_i^2/4\pi$ are given by
 $d\alpha_i^{-1}/(d\ln \mu) = -b_i/(2\pi)$,
 where $i=Y$, 2, and 3,
 and the coefficients of $U(1)_Y$, $SU(2)_L$, and $SU(3)_C$ gauge couplings are given by
 ($b_Y^{SM}$, $b_2^{SM}$, $b_3^{SM}$)=(41/6, $-19/6$, $-7$) in the SM.
For simplicity, we only consider a GUT normalization factor of 3/5 as in $SU(5)$ GUT,
 i.e. $b_1^{SM} = 41/10$.
Once particle contents in the model are fixed,
 contributions to $b_i$ are systematically calculated as in Table~\ref{bi}~\cite{Jones:1981we}.
In this table, fermions included vector-like to avoid the gauge anomalies.

\begin{table}[ht]
\begin{center}
\caption{Contributions to $b_i$ from fermions (left) and scalars (right).
$U(1)_Y$ hypercharge "$a$" can take different values for different representations,
 and an electric charge is given by $Q_{em} = I_3 + a/2$ with isospin $I_3$.
$b_1$ is given by $b_1=3/5 \times b_Y$.}
\begin{tabular}{|c|c|}\hline
Irreducible representation & Contribution to ($b_1$, $b_2$, $b_3$) \\
($SU(3)_C$, $SU(2)_L$, $U(1)_Y$) & by fermions\\
\hline \hline
(1, 1, 0) & (0, 0, 0)\\ \hline
(1, 1, $a$)$\oplus$(1, 1, $-a$) & ($\frac{1}{5}a^2$, 0, 0)\\ \hline
(1, 2, $a$)$\oplus$(1, 2, $-a$) & ($\frac{2}{5}a^2$, $\frac{2}{3}$, 0)\\ \hline
(1, 3, 0) & (0, $\frac{4}{3}$, 0)\\ \hline
(1, 3, $a$)$\oplus$(1, 3, $-a$) & ($\frac{3}{5}a^2$, $\frac{8}{3}$, 0)\\ \hline
(3, 1, $a$)$\oplus$($\overline{3}$, 1, $-a$) & ($\frac{3}{5}a^2$, 0, $\frac{2}{3}$)\\ \hline
(3, 2, $a$)$\oplus$($\overline{3}$, 2, $-a$) & ($\frac{6}{5}a^2$, 2, $\frac{4}{3}$)\\ \hline
(3, 3, $a$)$\oplus$($\overline{3}$, 3, $-a$) & ($\frac{9}{5}a^2$, 8, 2)\\ \hline
(6, 1, $a$)$\oplus$($\overline{6}$, 1, $-a$) & ($\frac{6}{5}a^2$, 0, $\frac{10}{3}$)\\ \hline
(6, 2, $a$)$\oplus$($\overline{6}$, 2, $-a$) & ($\frac{12}{5}a^2$, 4, $\frac{20}{3}$)\\ \hline
(6, 3, $a$)$\oplus$($\overline{6}$, 3, $-a$) & ($\frac{18}{5}a^2$, 16, 10)\\ \hline
(8, 1, 0) & (0, 0, 2)\\ \hline
(8, 1, $a$)$\oplus$(8 1, $-a$) & ($\frac{8}{5}a^2$, 0, 4)\\ \hline
(8, 2, $a$)$\oplus$(8, 2, $-a$) & ($\frac{16}{5}a^2$, $\frac{16}{3}$, 8)\\ \hline
(8, 3, 0) & ($\frac{12}{5}a^2$, $\frac{32}{3}$, 6)\\ \hline
(8, 3, $a$)$\oplus$(8, 3, $-a$) & ($\frac{24}{5}a^2$, $\frac{64}{3}$, 12)\\ \hline
\end{tabular} \hspace{0.5cm}
\begin{tabular}{|c|c|}\hline
Irreducible representation & Contribution to ($b_1$, $b_2$, $b_3$) \\
($SU(3)_C$, $SU(2)_L$, $U(1)_Y$) & by scalar particles\\
\hline \hline
(1, 1, $a$) & ($\frac{1}{20}a^2$, 0, 0)\\ \hline
(1, 2, $a$) & ($\frac{1}{10}a^2$, $\frac{1}{6}$, 0)\\ \hline
(1, 3, $a$) & ($\frac{3}{20}a^2$, $\frac{2}{3}$, 0)\\ \hline
(3, 1, $a$) & ($\frac{3}{20}a^2$, 0, $\frac{1}{6}$)\\ \hline
(3, 2, $a$) & ($\frac{3}{10}a^2$, $\frac{1}{2}$, $\frac{1}{3}$)\\ \hline
(3, 3, $a$) & ($\frac{9}{20}a^2$, 2, $\frac{1}{2}$)\\ \hline
(6, 1, $a$) & ($\frac{3}{10}a^2$, 0, $\frac{5}{6}$)\\ \hline
(6, 2, $a$) & ($\frac{3}{5}a^2$, 1, $\frac{5}{3}$)\\ \hline
(6, 3, $a$) & ($\frac{9}{10}a^2$, 4, $\frac{5}{2}$)\\ \hline
(8, 1, $a$) & ($\frac{2}{5}a^2$, 0, 1)\\ \hline
(8, 2, $a$) & ($\frac{4}{5}a^2$, $\frac{4}{3}$, 2)\\ \hline
(8, 3, $a$) & ($\frac{6}{5}a^2$, $\frac{16}{3}$, 3)\\ \hline
\end{tabular}
\label{bi}
\end{center}
\end{table}

\begin{figure}[ht]
  \begin{center}
          \includegraphics[clip, width=60mm]{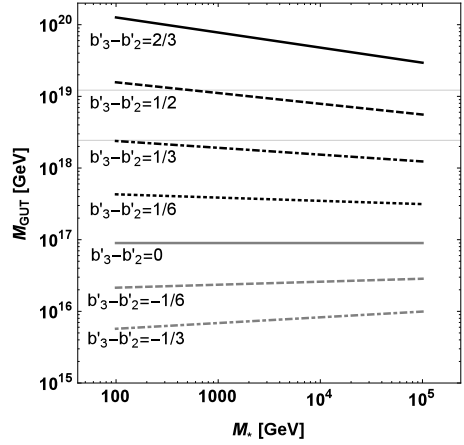}
  \end{center}
\caption{Relations between $M_*$ and $M_{GUT}$ for fixed $b'_3-b'_2$.
These lines correspond to $b'_3-b'_2=$ 2/3, 1/2, $\cdots$, and $-1/3$, respectively.
Two horizontal lines represent the Planck scale,
 i.e. $M_{Pl}=2.4\times10^{18}$\,GeV and $\sqrt{8\pi}M_{Pl}=1.2\times10^{19}$\,GeV, respectively.}
\label{GUTscale}
\end{figure}

To realize the GCU,
 we will consider extra particles with the TeV scale mass,
 which is motivated by avoiding the gauge hierarchy problem.
Using the solution of the one-loop RGEs,
 one can obtain the GCU conditions as
\begin{eqnarray}
	b'_i-b'_j = \frac{2\pi}{\ln\left(\frac{M_{GUT}}{M_*}\right)}
				\left(\alpha_i^{-1}(M_*)-\alpha_j^{-1}(M_*)\right)-(b_i^{SM}-b_j^{SM})
\label{b_dif},
\end{eqnarray}
 where $M_*$ and $M_{GUT}$ are the mass scale of extra particles
 and the GCU scale, respectively.
And, $b'_i$ are contributions of the extra particles as $b_i=b_i^{SM}+b'_i$.
From Table~\ref{bi},
 one can see $b'_3-b'_2 \propto 2/3$ and 1/6 for fermions and scalars, respectively.
Figure~\ref{GUTscale} shows relations between $M_*$ and $M_{GUT}$ for fixed $b'_3-b'_2$.

We can see that $M_{GUT}$ does not strongly depend on $M_*$
 once a value of $b'_3-b'_2$ is fixed.
It is worth noting that
 only $b'_3-b'_2=1/3$ or 1/2 can realize the GCU at the Planck scale,
 which are represented by two horizontal grid lines.\footnote{
If, however, we use two-loop RGEs and one-loop threshold corrections,
 values of gauge couplings in a high energy scale could have ${\cal O}(1)$ uncertainty.
Thus, the GCU could be realized at the Planck scale even for $b'_3-b'_2 = 1/6$ and 2/3.}
Thus, one can find that, when all extra particles are fermions,
 the GCU at the Planck scale cannot be realized.
This is the same result in Ref.~\cite{Giudice:2004tc}.
On the other hand,
 when extra particles include some relevant scalars such as (1, 2, $a$),
 the GCU can be realized at the Planck scale as follows.
For $b'_3-b'_2 = 1/3$, the GCU can be realized at $M_{GUT}\simeq2.0\times10^{18}\,{\rm GeV} \sim M_{Pl}$
 by extra particles satisfying
\begin{eqnarray}
	b'_3 = \frac{17}{6}+\frac{n}{6}\ (n=0,1,2,\cdots, {\rm and}\ 35),\qquad
	b'_2 = b'_3-\frac{1}{3},\qquad
	b'_1 \simeq b'_3-2.8,
\label{case1/3}
\end{eqnarray}
 where the minimum value of $b'_3$ is determined to satisfy $b'_1 \geq 0$,
 and the largest value of $n$ is determined to avoid the Landau pole.
In the same way,
 for $b'_3-b'_2 = 1/2$, the GCU can be realized at $M_{GUT}\simeq \sqrt{8\pi}M_{Pl}$
 by extra particles satisfying
\begin{eqnarray}
	b'_3 = \frac{10}{3}+\frac{n}{6}\ (n=0,1,2,\cdots, {\rm and}\ 33),\qquad
	b'_2 = b'_3-\frac{1}{2},\qquad
	b'_1 \simeq b'_3-3.2.
\label{case1/2}
\end{eqnarray}

\section{Realization of the GCU at the Planck scale} \label{sec:realization}

According to the above discussions,
 we systematically investigate possibilities of the realization of GCU at the Planck scale,
 and find that a number of combinations of extra particles satisfy Eq.\,(\ref{case1/3}) or (\ref{case1/2}).
For example,
 when we consider extra scalars are two $SU(2)_L$ doublets (1, 2, 0),
 the GCU can be realized around $M_{Pl}$
 by extra fermions shown in Table~\ref{combination}.
For simplicity,
 representation of extra fermions are the same as the SM fermions (with vector-like partners)
 and an $SU(2)_L$ adjoint fermion denoted by $W$.\footnote{
Stable TeV-scale particles with fractional electric charge (such as the $SU(2)_L$ doublet scalar (1, 2, 0))
 might cause cosmological problems.
In order to avoid the problems,
 the reheating temperature $T_R$ after the inflation should be about 40 times lower than
 the particle masses~\cite{Chung:1998rq},
 that is, $T_R \sim {\cal O}(10)\,{\rm GeV}$ in our cases.}

\begin{table}[ht]
\begin{center}
\caption{Examples of extra fermions,
 which satisfy Eq.\,(\ref{case1/3}) with two $SU(2)_L$ doublets (1, 2, 0).
In all cases, extra particle masses are $M_*=1$\,TeV,
 and the GCU is realized around $M_{Pl}$.
In the rightmost column, $n$ is given in Eq.\,(\ref{case1/3}).
}
\begin{tabular}{|l||l|c|c|}\hline
Extra fermions & ($b'_1$, $b'_2$, $b'_3$) & $\alpha_{GUT}^{-1}$ & $n$\\
\hline \hline
$Q\overline{Q} \times 1$ $\oplus$ $D\overline{D} \times 4$ $\oplus$ $W \times 1$ & ($\frac{6}{5}$, $\frac{10}{3}$, 4) & 28.0 & 7\\ \hline
$Q\overline{Q} \times 2$ $\oplus$ $D\overline{D} \times 3$ $\oplus$ $E\overline{E} \times 1$ & ($\frac{28}{15}$, 4, $\frac{14}{3}$) & 24.3 & 11\\ \hline
$Q\overline{Q} \times 2$ $\oplus$ $U\overline{U} \times 1$ $\oplus$ $D\overline{D} \times 2$ & ($\frac{28}{15}$, 4, $\frac{14}{3}$) & 24.3 & 11\\ \hline
$Q\overline{Q} \times 2$ $\oplus$ $D\overline{D} \times 4$ $\oplus$ $L\overline{L} \times 1$ $\oplus$ $E\overline{E} \times 1$ & ($\frac{38}{15}$, $\frac{14}{3}$, $\frac{16}{3}$) & 20.5 & 15\\ \hline
$Q\overline{Q} \times 2$ $\oplus$ $U\overline{U} \times 1$ $\oplus$ $D\overline{D} \times 3$ $\oplus$ $L\overline{L} \times 1$ & ($\frac{38}{15}$, $\frac{14}{3}$, $\frac{16}{3}$) & 20.5 & 15\\ \hline
$Q\overline{Q} \times 2$ $\oplus$ $U\overline{U} \times 2$ $\oplus$ $D\overline{D} \times 3$ $\oplus$ $W \times 1$ & ($\frac{16}{5}$, $\frac{16}{3}$, 6) & 16.8 & 19\\ \hline
$Q\overline{Q} \times 3$ $\oplus$ $U\overline{U} \times 2$ $\oplus$ $D\overline{D} \times 2$ $\oplus$ $E\overline{E} \times 1$ & ($\frac{58}{15}$, 6, $\frac{20}{3}$) & 13.1 & 23\\ \hline
$Q\overline{Q} \times 3$ $\oplus$ $U\overline{U} \times 3$ $\oplus$ $D\overline{D} \times 1$ & ($\frac{58}{15}$, 6, $\frac{20}{3}$) & 13.1 & 23\\ \hline
\end{tabular}
\label{combination}
\end{center}
\end{table}

Next, we consider extra fermions having different masses,
 which are taken as $0.5\,{\rm TeV} \leq M \leq 10\,{\rm TeV}$.
Actually, we take only lepton masses 0.5\,TeV,
 since lower bounds of vector-like lepton and quark masses are around 200\,GeV and 800\,GeV, respectively
 ~\cite{CMS:2012ra, Chatrchyan:2013uxa,Aad:2014efa}.
In Table~\ref{combination_f}, we show
 extra fermions which realize the GCU around $\sqrt{8\pi}M_{Pl}$.
In the table, "$W\times 1$ (0.5)" shows 
 one (1, 3, 0) fermions with a mass of 0.5\,TeV, and so on.

\begin{table}[ht]
\begin{center}
\caption{Examples of extra fermions
 which realize the GCU around $\sqrt{8\pi}M_{Pl}$.
In the leftmost column, the values in brackets show the corresponding fermion masses with a unit of TeV.
}
\begin{tabular}{|p{295pt}||l|c|}\hline
Extra fermions & ($b'_1$, $b'_2$, $b'_3$) & $\alpha_{GUT}^{-1}$ \\
\hline \hline
$W\times 1$ (0.5) $\oplus$ $U\overline{U}\times 1$ (1) $\oplus$ $Q\overline{Q}\times 2$ (10) $\oplus$ $D\overline{D}\times 4$ (10) & ($\frac{12}{5}$, $\frac{16}{3}$, 6) & 19.1 \\ \hline
$E\overline{E}\times 2$ (0.5) $\oplus$ $Q\overline{Q}\times 2$ (2) $\oplus$ $Q\overline{Q}\times 2$ (10) $\oplus$ $D\overline{D}\times 4$ (10) & ($\frac{46}{15}$, 6, $\frac{20}{3}$) & 14.9 \\ \hline
$L\overline{L}\times 1$ (0.5) $\oplus$ $E\overline{E}\times 1$ (0.5) $\oplus$ $Q\overline{Q}\times 1$ (1) $\oplus$ $U\overline{U}\times 1$ (1) $\oplus$ $Q\overline{Q}\times 2$ (10) $\oplus$ $D\overline{D}\times 4$ (10) & ($\frac{56}{15}$, $\frac{20}{3}$, $\frac{22}{3}$) & 11.1 \\ \hline
$E\overline{E}\times 1$ (0.5) $\oplus$ $W\times 1$ (0.5) $\oplus$ $U\overline{U}\times 2$ (4) $\oplus$ $Q\overline{Q}\times 3$ (10) $\oplus$ $D\overline{D}\times 4$ (10) & ($\frac{22}{5}$, $\frac{22}{3}$, 8) & 7.95 \\ \hline
\end{tabular}
\label{combination_f}
\end{center}
\end{table}

\begin{figure}[ht]
  \begin{center}
          \includegraphics[clip, width=70mm]{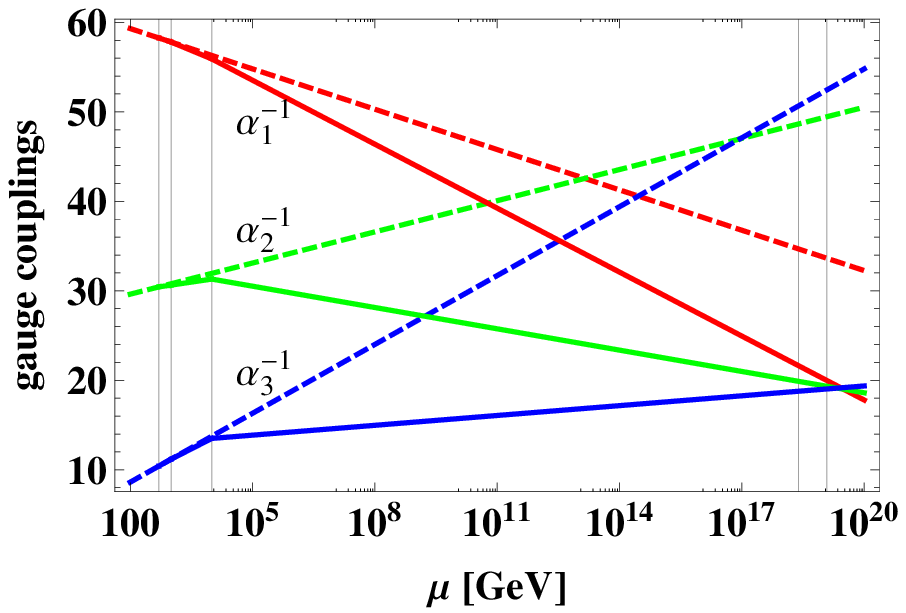} \hspace{10mm}
          \includegraphics[clip, width=70mm]{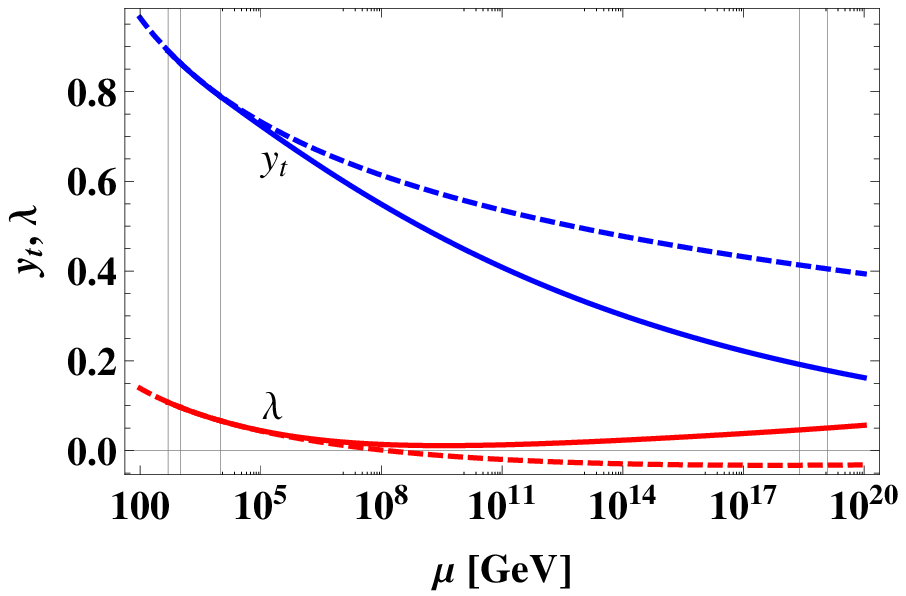}
  \end{center}
\caption{Runnings of the gauge couplings (left),
 and the top Yukawa and the Higgs quartic couplings (right).
These figures correspond to the first one of Table~\ref{combination_f}.
The solid and dashed lines correspond to the extended SM and the SM, respectively.
Three vertical lines represent
 0.5\,TeV, 3\,TeV, 10\,TeV, $M_{Pl}$, and $\sqrt{8\pi}M_{Pl}$, respectively.}
\label{running_f}
\end{figure}

In Fig.\,\ref{running_f}
 we show runnings of the gauge, the top Yukawa, and the Higgs quartic couplings,
 which correspond to the first one of Table~\ref{combination_f}.
Here, we assume extra fermions do not strongly couple with the Higgs doublet,
 and then significantly change runnings of the top Yukawa and the Higgs quartic couplings.
Since $\beta$-functions of the gauge couplings change several times,
 our previous naive analyses are modified,
 that the GCU can be realized around $\sqrt{8\pi}M_{Pl}$
 by extra fermions satisfying $b'_3-b'_2=2/3$.
To realize the GCU,
 all the gauge couplings are large compared to those in the SM,
 which lead the smaller $y_t$.
Then, $\lambda$ becomes larger and remains in positive value up to the Planck scale,
 since both the smaller $y_t$ and the larger $g_i$ make $\beta_\lambda$ become larger.
Thus, when the GCU is realized at the Planck scale,
 we expect the vacuum can become stable.

\section{Summary} \label{sec:summary}

We have investigated possibilities of the GCU at the Planck scale
 in the extended SM which includes extra particles around the TeV scale.
We have found that
 the GCU at the Planck scale can be realized
 when extra particles include some relevant scalars,
 while it cannot be realized (up to one-loop level)
 when all extra particles are fermions and their masses are the same.
On the other hand,
 when extra fermions have different masses,
 the GCU around $\sqrt{8\pi}M_{Pl}$ can be realized.
In this case, the vacuum can become stable
 because of the change of the running gauge couplings.

\begin{acknowledgments}
We thank N. Haba, R. Takahashi and H. Ishida for useful discussion and fruitful collaborations.
The works of Y.Y. is supported
 by Research Fellowships of the Japan Society for the Promotion of Science for Young Scientists,
 Grants No. 26$\cdot$2428.
\end{acknowledgments}

\bigskip 


\end{document}